\documentclass[12pt]{article}

\newcommand\snoska[1]{\footnote{#1}}
\renewcommand\snoska[1]{}

\newcommand{\bec}[1]{\mbox{\boldmath $ #1$}}
\newcommand{\sun}{\ensuremath{\odot}} 
\newcommand{\grad}{{\bf \nabla\,}}
\newcommand{\curl}{{\rm curl}\,}

\begin{document}

\righthyphenmin=4
\lefthyphenmin=4

\sloppy

\baselineskip=0.84cm


\centerline{\bf {ESTIMATES OF CURRENT HELICITY AND TILT OF
}}
\centerline{\bf {
SOLAR ACTIVE REGIONS AND JOY'S LAW
}}
\vspace{0.1cm}
\vspace{0.1cm}

\centerline{{K.~Kuzanyan$^{1,2}$, N.~Kleeorin$^{3,4}$,
I.~Rogachevskii$^{3,4}$, D.~Sokoloff$^{5,6}$, H.~Zhang$^{1}$}
}
\vspace{0.1cm}
\centerline{
 $^{1}$National Astronomical Observatories, Chinese Academy of Sciences, 
}
\centerline{
Key Laboratory for Solar Activity, Beijing 100012, China
}
\centerline{
 $^{2}$IZMIRAN, Russian Academy of
Sciences, Moscow, Troitsk, Kaluzhskoe sh.4, 108840,
Russia
}
\vspace{0.1cm}
\centerline{
$^{3}$Department of Mechanical Engineering, Ben-Gurion University of Negev
}
\centerline{
POB 653, 84105 Beer-Sheva, Israel}
\vspace{0.1cm}
\centerline{
 $^{4}$Nordita, KTH Royal Institute of Technology and Stockholm University,
        Stockholm, Sweden}
\vspace{0.1cm}
\centerline{
 $^{5}$Department of Physics, Moscow University, 119992 Moscow,
Russia
}
\vspace{0.1cm}
\centerline{
 $^{6}$Moscow Center of Fundamental and Applied Mathematics, Moscow, 119991, 
Russia}

\centerline{Received  March 1, 2020}

\centerline{Revised April 30, 2020}
\centerline{Accepted  .....2020}

\noindent

\vspace{0.2cm}



\begin{abstract}
{\baselineskip=0.84cm
The tilt angle, current helicity and twist of solar magnetic fields can be observed in solar active regions. We carried out estimates of these parameters by two ways. Firstly, we consider the model of turbulent convective cells (super-granules) which have a loop floating structure towards the surface of the Sun. Their helical properties are attained during the rising process in the rotating stratified convective zone. The other estimate is obtained from a simple mean-field dynamo model that accounts magnetic helicity conservation. The both values are shown to be capable to give important contributions to the observable tilt, helicity and twist.
}

\end{abstract}

\section{Introduction}

An important property of the solar magnetic
fields in sunspots and active regions is their
helical nature. Quantitative studies of this
effect have been carried out for a long time,
e.g., Seehafer (1990), Pevtsov \& Canfield
(1994), Bao \& Zhang (1998), Hagino \& Sakurai
(2004). \cite{Zhang_ea2010}(thereafter Z10) presented a systematic
study of
current helicity density and
twist of photospheric magnetic fields in solar
active regions. The obtained butterfly diagram 2D
(latitude-time) plot covers the two consequent
solar cycles. The quantities which are presented
in Z10 are the radial part of current helicity
and twist (force-free factor) averaged over a
statistically significant sample of active
regions. Vector magnetographic data have been
collected for more than 6,000 individual
magnetograms over the period of almost two
sunspots cycles 1988-2005, and they demonstrate
significant cyclic dynamics of these quantities.
At the same time they give relatively reliable
magnitudes of these helical quantities which
would be attractive to obtain by some ad-hoc
analysis of the very basic process of sunspot
formation and compare it with the observational
values.

The magnetic helicity plays an importance role in
hydromagnetic dynamo theory as it is an inviscid
invariant which is conserved in turbulent
convective motion for very large magnetic
Reynolds numbers. It can be used as a constraint
for the theoretical modelling. Current helicity
and twist are observational tracers of magnetic
helicity. Furthermore, as the observational
results have been complimented by constructions
of dynamo models based on helicity conservation
argument (e.g. Kleeorin et al. 2003; Zhang et al.
2012, and references therein).
In the present paper we estimate the tilt (the angle 
between a line connecting the leading and following sunspots 
and the solar equator), current helicity and twist of magnetic fields in solar active regions.

\section{The Model}

Let us consider a simple bipolar active
region with the distance $L$ between the opposite
polarities. We use a model of a turbulent convective cell of the size of
super-granulation with a depth $L/2 \approx 10^9{\rm cm}$.
This scale is of the order of the density stratification scale in the convective zone $L/2 \approx
H_\rho=-\displaystyle{\left[{d\over
dr}{\log{\rho_0(r)}}\right]^{-1}}$, where
$\rho_0$ is the fluid density at the depth $H=10^9{\rm cm}$.
This scale we associate with the depth of the sunspot formation region.
In a classical Rayleigh-B\'{e}nard convective roll, horizontal and vertical scales are the same.
A superposition of the three convective rolls forms a hexagonal structure (see, e.g., \cite{Chandra1961}, Chap.~16, p.~48, Fig.~7a). This implies that the ratio of horizonal to vertical sizes in the hexagonal structure is about 2.
In the Sun and stars the convection is fully turbulent and is different from classical Rayleigh-B\'{e}nard convection. In particular, the convection rolls are formed with their horizontal scales greater than the vertical scales approximately by factor 2. Similar phenomena are observed in Earth's atmosphere and other natural turbulent convection systems.
Large-scale structures like the convection rolls can be isolated from turbulent eddies using scale separation ideas (Elperin et al 2002; Bukai et al. 2009) being applied to solar active regions.
This consideration is in agreement with observations of super-granulation convection in the Sun.
Therefore, the total horizontal extent of this active region is $2L$.
Let us assume that the active region scales are small
compared to the solar radius, and put it at
heliographic latitude $\phi$ being counted from
the solar equator. In terms of observable active
regions, we estimate this scale by order of
magnitude as $L \sim 20-50 \,{\rm Mm} =
(2-5)\cdot10^{9}\,{\rm cm}$.

We consider the momentum equation for ${\bf u}$
applying the anelastic approximation at the boundary between the solar convective
zone and the photosphere
\begin{equation}
\frac{\partial {\bf u}}{\partial t} = -\grad \left(
\frac{p_{tot}}{\rho_0}\right)
-{\bf g} S
 + \mathcal{{\bf F}}_{mag}
 + \mathcal{{\bf F}}_{hd}
 + \mathcal{{\bf F}}_{visc}
 + \mathcal{{\bf F}}_{cor} ,
\label{vandak}
\end{equation}
which includes the effects of the total pressure, hydrodynamic and magnetic buoyancy,
nonlinear local hydrodynamic Coriolis force, viscous forces and global Coriolis force.
The formulation of equation (\ref{vandak})originates from the book \cite{vand76}, also see the recent work of \cite{Kleeorin2020}. 
The detailed forms of these quantities are decribed in Appendix~A.

Let us estimate sunspot twisting time $\tau_D$ as the ratio
$L/v_a$, where Alfven speed is
$v_a=B_{eq}/ \sqrt{4\pi
\rho_0}$. At the upper part of the convective zone typical equipartition value of
the mean magnetic field is $B_{eq} \sim 300\, {\rm G}$, and
density of the solar plasma $\rho_0$
according to estimates of Spruit (1974),
is of the order of
$4.5 \cdot 10^{-7} {\rm g}\,{\rm cm}^{-3}$.
Thus, Alfven speed is of the order of
$v_a \sim 1.2 \cdot 10^5 {\rm cm}\,{\rm s}^{-1}$,
and therefore the time scale
$\tau_D \sim (2-4) \cdot 10^4{\rm s} \approx 6-12$~hours.

The duration of the flux tube emerging from the bottom of the convective cell
is of the order of $\tau_F=L/2u_r$.
We use the anelastic approximation for convection in the solar convection zone ${\rm div} (\rho \bec u) = 0$. The density stratification is in the radial direction, so we have ${\rm div} \bec u =
-u_r\displaystyle{\frac{d}{dr}} \log \rho
 = \displaystyle{\frac{u_r}{H_\rho}}
\approx \displaystyle{\frac{1}{\tau_F}}$, where $H_\rho$ is density stratification scale.
Estimate for the radial component of vorticity $\omega_r$ for the motion of bipolar sunspots
is obtained in Appendix~A:
\begin{equation}
\omega_r
\approx
- \Omega_\odot \displaystyle{\frac{\tau_D}{\tau_F}}
\left(
4\,  \sin{\phi}
+
\frac{1}{\xi}
{\cos{\phi}}%
\right)
\,,
\end{equation}
where instead of co-latitude $\theta$ we use the latitude $\phi=\pi/2-\theta$,
then $\tau_F=H_\rho/u_r$ 
and $\xi$ is defined below.
The latitudinal derivative of the vertical convective velocity can be estimated as
$\displaystyle{\frac{1}{r}\frac{\partial u_r}{\partial \theta}}
\approx
\displaystyle{\frac{u_r}{\xi L}}$, where $\xi$ being dependent on the structure of active regions varies in sign, and in absolute value is around $1-2$ due to the hexagonal structure of convection.
The sign of $\xi$ can be considered random for super-granulation convection in the Sun.
If we consider roll-like convection, it is close to unity while for hexagonal convective cells comprising of three rolls, it is close to 2 (see Elperin et al 2002; Bukai et al. 2009).

\subsection{Estimate for Tilt of Sunspots}

Given the time of evolution of the active region
during the sunspot formation (which contributes to
the tilt angle of the opposite polarities), is
comparable with the flux tube emerging time
$\tau_F$, so that we can estimate the tilt as
\begin{equation}
 \delta \approx w_r \tau_F
= \displaystyle{-\frac{2\pi\tau_D}{T_{\odot}}}
\left(4\sin{\phi}
+\displaystyle{\frac{\cos{\phi}}{\xi}}
\right) ,
 \label{tilt}
\end{equation}
where the solar siderial rotation period ${T_{\odot}}\approx 25\,$d
approximately corresponding to Carrington rotation.
The coefficient in the front of the brackets in this formula is of order $0.25-0.5$.

Since the value of $\sin{\phi}$ for low
latitudes where sunspots mainly occur is comparable with the latter term within the brackets, that is of order $1/8-1/4$, we can expect the tilt angles of individual active regions to vary a lot and even change sign. This estimate also implies the less variability in tilt angles the higher latitude, which may need to be verified with observations.
Given the sign of $\xi$ randomly fluctuates and its value varies with super-granulation, for the averaged tilt $\langle \delta \rangle $ we may obtain the range $(0.25-0.5) \sin{\phi}$, which gives the order
up to $15^\circ$ for middle latitudes, and so it fits perfectly well with the observational results of Howard (1991).
The observational magnitude of tilt is indeed
increasing with departure from the solar equator
almost linearly with latitude, or like
$\sin{\phi}$, and for middle-range latitudes it is
in average $5-15^\circ$, see e.g. Stenflo \& Kosovichev (2012);
\cite{TlatovIllarionov2013,StenfloKosovichev2012} and references therein.

\subsection{Estimates for Current Helicity and Twist from Solar Observations}

We estimate the current helicity in terms of
the magnetic energy and the vertical variation scale of
magnetic field. This scale is of the order
of several density stratification
scales $H_\rho$.
The flux tubes are mainly formed by large-scale MHD instability
(e.g., the magnetic buoyancy instability 
\cite{1955ApJ...121..491Parker_sunspot}
and the negative effective magnetic pressure instability \cite{Kleeorin1990,Brandenburg2016})
from the dynamo generated large-scale magnetic field.
These magnetic flux tubes are
unstable when the mean field strength exceeds a critical
value. The rise of magnetic flux tubes to the surface
due to magnetic buoyancy can cause formation of sunspots and active regions.
As the horizontal fields $B_x$ and $B_y$ are getting
tilted with the active region formation process,
they form horizontal gradients which are
observable in high resolution vector
magnetographic measurements and used for
computation of electric currents, and,
subsequently, current helicity density. These
titled distortions can be estimated as
\begin{equation}
 \Delta B_y \approx
\delta
\displaystyle{
\frac{\partial B_x}{\partial x}} x
\qquad
{\rm and}
\qquad
 \Delta B_x \approx
-\delta
\displaystyle{
\frac{\partial B_y}{\partial y}} y .
\end{equation}
Using solenoidality condition ${\rm div} \bec B = 0$,
we determine the vertical component of the $\curl$ as
\begin{equation}
(\curl B)_z =
\displaystyle{
\frac{\partial \Delta B_y}{\partial x}}
-
\displaystyle{
\frac{\partial \Delta B_x}{\partial y}}
\,=\,
\delta
\displaystyle{
\left(
\frac{\partial B_x}{\partial x}
+\frac{\partial B_y}{\partial y}
\right)
}
=-\delta
\displaystyle{
\frac{\partial B_z}{\partial z}
}
\,.
\end{equation}
Now we can relate the vertical part of the current helicity with twist as
\begin{equation}
H_c=B_z(\curl B)_z =
-\delta
B_z\displaystyle{
\frac{\partial B_z}{\partial z}
}=
\delta\,
\displaystyle{
\frac{B_z^2}{H_B}
}
\,,
\label{heli2}
\end{equation}
where $H_B$ is the vertical magnetic field variation scale, which is of the order of
the sunspot size ($\sim 10-20$ Mm $= 1-2\cdot10^9$ cm).

Correspondingly, the estimate of twist $\Upsilon$ for typical tilt angles of order $\delta \sim 0.1-0.2$ ($5-10^\circ$) reads
\begin{equation}
\Upsilon=
\delta/H_B
\sim 10^{-10}\,{\rm cm}^{-1} = 10^{-8}\,{\rm m}^{-1}
\,,
\label{twist}
\end{equation}
which matches quite well the order of magnitude of the observational results, e.g. \cite{Zhang_ea2002}; Zhang el al. (2010).
Notice that the vertical magnetic field energy proportional to ${B_z^2}$ is of the order of the equipartition magnetic field.

For estimation of current helicity we take into account that the typical value of the magnetic field varies from a few G in the quiet Sun to kG in sunspots. Keeping in mind, however, that for comparison with vector magnetographic observations we must focus mainly on penumbral parts of active regions, the reasonable estimate for the field would be hundreds of gauss. Setting $B_z \sim 300$~G we obtain the estimate for $H_c \sim 10^{-3} {\rm G}^2 {\rm m}^{-1}$, in agreement with observations of Bao \& Zhang (1998); Zhang el al. (2010).
These naive add-hoc estimates of the tilt angles as well as current helicity and twist in solar active regions based on local considerations are in accordance with the observational ranges of current helicity and twist.

\section{Estimates for Current Helicity and Twist in Dynamo Models}

After we have estimated the tilt angles as well as current helicity and twist using local effects for rising flux tubes in solar active regions, we are going to estimate these values from the axially symmetric spherical shell dynamo model by Zhang et al. (2012), see also references therein.
Let us use Equation (5) of Zhang et al. (2012) for the mean current helicity
of the active region,
\begin{eqnarray}
H_c^{\rm AR}=
\langle {\bf B}^{\rm AR} {\bf
\cdot} \bec{\rm curl} \,{\bf B}^{\rm AR} \rangle
\approx
- {1 \over L^2_{\rm ar}} {\bec{ A}} {\bf \cdot} {\bec{ B}}
= - {B_{*}^2 R_\odot \over
L^2_{\rm ar}} \tilde{A} \tilde{B},
 \label{D1}
\end{eqnarray}
where $A=R_\sun \tilde A B_{*}$, 
$B=\tilde B B_{*}$; $B_{*}=10 \rho_0^{1/2} \eta_T/R_\sun$ is the characteristic magnetic field produced by the dynamo mechanism and $\eta_T$ is the turbulent magnetic diffusion coefficient 
(see Zhang et al. (2012); Kleeorin et al. (1995)), and
$R_\sun \approx 7\times10^{10}\,{\rm cm}$ is the solar radius.
The twist of magnetic fields of an active region can be estimated as
\begin{eqnarray}
\Upsilon \equiv {\langle {\bec B}^{\rm AR} {\bec
\cdot} \bec{\curl} \,{\bec B}^{\rm AR} \rangle
\over \langle ({\bec B}^{\rm AR})^2 \rangle}
\approx - {{\bec{ A}} {\bec \cdot} {\bec{ B}}
\over L^2_{\rm AR} {\bec{ B}}^2} \approx
-{{\tilde{A}} \over {\tilde{B}}}
{{R_\odot} \over{ L^2_{\rm aAR}}},
 \label{D2}
\end{eqnarray}
where we assumed that
$\langle ({\bf B}^{\rm AR})^2 \rangle \approx {\bec{ B}}^2$.
Note that the product and ratio of $\tilde A$ and $\tilde B$ vary within the solar cycle but does not vary from one (odd/even) cycle to another (even/odd). The sign of ${\bec{A}} \cdot {\bec{B}}$ is mainly negative (~3/4 of the period), so the sign of $\Upsilon >0$, i.e. opposite to the one produced by Coriolis force.
According to observations the horizontal size of the active region is about the size of a super-granule,
i.e., $L_{\rm AR} \sim (2-5)\cdot H_\rho$,
where $H_\rho \sim 10^9$~cm.
So, we estimate $L_{AR} \sim (2-5)\cdot 10^{9}\,{\rm cm}
= 20-50\,{\rm Mm} $.

Using Eq.~(\ref{D2}), we estimate the twist of the magnetic field in the active region as
\begin{eqnarray}
\Upsilon \approx -
{{\tilde{A}} \over {\tilde{B}}}
{{R_\odot} \over
{{L_{\rm AR}}^2}} \sim
 - (0.3-1) \cdot 10^{-10}
{\rm cm}^{-1},
 \label{D3}
\end{eqnarray}
where typically $\tilde{A}/\tilde{B} \sim
6\times 10^{-3}$
(see Appendix~B and also the figures of \cite{Zhang et al. (2012)}).
This value is of order $10^{-2}$ which is typical for the most kinematic dynamo models of $\alpha\Omega$ type.
Now, let us take the value of $\tilde{B}$ of
order $0.2-0.5$ (as it is on in the bottom of the
convective zone). If we adopt the value of
$B_{eq}\sim 500-1000 \,{\rm G}$ as somewhere in the
sunspot umbra near or just beneath the photosphere,
we estimate the magnitude for the current helicity as
\begin{eqnarray}
H_c^{\rm ar} = \Upsilon {\bec{ B}}^2 {\tilde{B}}^2
 \sim -  10^{-5}
{\rm G}^{2} {\rm cm}^{-1}
= -  10^{-3} {\rm G}^{2} {\rm m}^{-1}
 \,.
 \label{D4}
\end{eqnarray}
This value is comparable with the observational
results of Zhang et al. (2010) giving the order
of $(1-2)\cdot10^{-5}\,{\rm G}^2\,{\rm cm}^{-1}$
for current helicity, and
$(1-2)\cdot10^{-10}\,{\rm cm}^{-1}$ for the
twist.

We can apply the other method of estimation of ratio $A/B$ using the numerical simulations of the 2D mean-field dynamo model in a spherical shell, e.g., by Zhang et al. (2012). For the variety of depths and latitudes, this ratio is of order $10^{-3}-10^{-2}$, most typically $(2-7)\cdot10^{-3}$ (private communications with late David Moss
\footnote{deceased in 2020}).
The estimate~({\ref{D3}}) for twist yields the same order of magnitude $ -(10^{-10}- 10^{-11}){\rm cm}^{-1}$ as estimated above.
For small-size active regions, $L \sim (0.5-1)\cdot H_\rho$, the contributions to current helicity from both
effects caused by the Coriolis force and the dynamo mechanism are getting comparable.

\section{Discussion}

In this paper we estimated the effects of two mechanisms of formation of tilt, twist and current helicity in solar active regions.
One mechanism is related to the action of Coriolis force on rising active regions, so for that  
the original magnetic field is not assumed twisted at the initial stage of rising magnetic flux tube.

The second mechanism is related to production of magnetic helicity by the dynamo process on the mean magnetic field. The study by Zhang et al. (2012) shows that the rising magnetic fields may already be helical at the very beginning stage because the magnetic helicity is produced together with the mean magnetic field generation.
The magnetic helicity of the mean field has the opposite sign to the current helicity of the active region.
The second mechanism of the origin of tilt, twist and current helicity is apparently independent of the first one.

These two mechanism cause opposite sign contributions to the tilt, twist and current helicity in the most but not all phases of the solar cycle.
The overall sign of tilt, twist and current helicity may depend on the particular ratio of the two contributions at the given time and location.
This may explain certain irregularities of the tilt from the classic hemispheric rule (e.g. \cite{TlatovIllarionov2013,Pevtsov_ea_rev2014}).

Having estimated the tilt, current helicity and twist from local considerations in active regions we notice that even the sign of them may vary from one active region to another.
On the other hand,
the averaging of the tilt, current helicity and twist
over all active regions does not depend on the phase of solar cycle and the sign of magnetic field (Hale's polarity law), so from one cycle to another cycle the sign of these quantities holds.
This is in good agreement with Joy's law for tilt as well as Hemispheric Sign Rule for current helicity and twist. Given the considered effects do not vary much with the phase of the solar cycle, we may have the same trend everywhere. However, as the quantities which are produced by the solar dynamo may vary with the phase of the cycle, we can impute this fact to systematic change in sign and reversal of the Hemispheric Sign Rule for current helicity and twist noticed by Bao et al. (2000) and obtained in detail for some phases of the solar cycle and some latitudes by Z10.

The results of our estimations show that the
value of helicity which can be formed by flux
tube arising in the convective zone during the
sunspot formation can well be comparable
with amount of helicity generated by the entire
dynamo process. This makes important simultaneous
studies of the origin of helicity in the Sun by
both mechanisms 
(see, \cite{Kleeorin2020}).

Our vision of the phenomena of helicity and tilt in solar active regions is significantly different from one reported by Longcope et al. (1998), see also Fisher, Fan et al. (1998), as the contribution to helicity and tilt due to Coriolis force is linearly proportional to the value of tilt while in those papers it appears quadratic (with the sign oppisite to the sign of tilt) in the value of tilt. In our vision, we operate not with helicity of magnetic tubes but with helicity of physical magnetic field in active regions. This naive consideration looks to us more relevant to observations. We impute the variability of sign of overall twist and helicity by the evolution of the dynamo contribution with time-latitudinal evolution in the solar cycle on the one hand, and the relatively stable contribution by Coriolis force, on the other.

However, these two mechanisms look apparently independent only from the first instance. The reason is that the current helicity is intimately related to magnetic helicity (see, e.g. \cite{BergerField1984}; \cite{Pevtsov_ea_rev2014} and references therein) and in absence of helicity flux the latter is a local integral in a non-dissipative MHD flow. In the scales which are much greater than turbulent scale of basic granulation and less than the typical size of an active region, we expect that the current helicity of the active region is determined by the magnetic helicity produced by the mean-field dynamo mechanism, see formula (\ref{D1}). 

%
%
In the presence of helicity fluxes some fraction of the dynamo-born helicity along with some (let us assume the same fraction) of helicity due to Coriolis force are removed from the 
photosphere and injected into the solar corona. Let us denote the fraction of this helicity injection as $\epsilon$. Then combining formulae (\ref{tilt}), (\ref{D1}), and (\ref{D4}), we can derive the following expression for the remaining total current helicity of an active region

\begin{eqnarray}
H_c^{\rm ar TOT} =
-\, \epsilon
\displaystyle{\frac{2\pi\tau_D}{T_{\odot}}}
\left(
4\,\sin{\phi}
+
\frac{1}{\xi}
{\cos{\phi}}
\right)
\frac{B_z^2}{H_B}
-\, (1-\epsilon)
{1 \over L^2_{\rm ar}} {\bec{ A}} {\bf \cdot} {\bec{ B}}  \,.
 \label{D5}
\end{eqnarray}
Hereby the ejection of helicity from an active region into the corona is

\begin{eqnarray}
H_c^{\rm ar FL} = 
 \epsilon
\displaystyle{\frac{2\pi\tau_D}{T_{\odot}}}
\left(
4\,\sin{\phi}
+
\frac{1}{\xi}
{\cos{\phi}}
\right)
\frac{B_z^2}{H_B}
-\, \epsilon
{1 \over L^2_{\rm ar}} {\bec{ A}} {\bf \cdot} {\bec{ B}}  \,.
 \label{D6}
\end{eqnarray}
Note that the sum of the total remaining and the flux parts of helicities in formulae (\ref{D4}) and (\ref{D5}) is equal to the amount of helicity produced by dynamo as in formula (\ref{D1}). The particular value of $\epsilon$ in formula (\ref{D5}) is not known and it can be estimated from comparison of theoretical dynamo models and the observational data for tilt, vertical magnetic field, twist and current helicity in solar active regions. The main contribution in the total helicity for large active regions is probably due to Coriolis force while for smaller one from dynamo generation mechanism. The distinction between the two can be determined by the latitude and phase of the solar cycle. This question requires further investigation with the use of calibrated dynamo models and available observational data. Please also note that the first part (contribution from large active regions) is scaled by $\tau_D \sim L$, so the relatively to the second part it is scaled as $\sim L^3$.
If for equation (\ref{D5}) we divide both sides by $B_z^2$ then we can treat this formula as expression for overall effective tilt. It contains both constant and oscillatory with the cycle parts.

\vspace{0.1cm}
\noindent
{\bf Acknowledgements.}

Stimulating discussions with participants
of the NORDITA program on ''Solar Helicities in Theory and Observations:
Implications for Space Weather and Dynamo Theory"
(March 2019) are acknowledged. NK, KK, IR would like to acknowledge the hospitality of NORDITA for this and earlier occasions where this work has been outlined and developed. Private communications with late David Moss are acknowledged.
The work of KK concerning the tilt and twist computations was supported by the grant from the Russian Science Foundation
(RNF 18-12-00131) at the Crimean Astrophysical Observatory.
DS would like to acknowledge support from RFBR grant 18-02-00085.
The observational estimates of helicity and twist are obtained due to collaboration of IZMIRAN team with National Astronomical Observatories of China, Key Laboratory for Solar Astivity, Chinese Academy of Sciences, supported in part by CAS PIFI visiting program and RFBR of Russia - NNSF of China joint grant 19-52-53045 GFEN.a.
HZ would also like to acknowledge support by grants from the National Natural Science Foundation of China (NSFC 11673033, 11427803, 11427901) and by Huairou Solar Observing Station, Chinese Academy of Sciences.

\newpage

\appendix

\section{Momentum equation in relaxation approximation}

We use the momentum equation
applying the anelastic approximation at the boundary between the solar convective
zone and the photosphere.
The terms appeared in equation (\ref{vandak}) are
${\bf w}=\curl {\bf u}$ for the vorticity,
$\displaystyle{p_{tot}=p + \frac{\rho {\bf u}^2}{2} + \frac{{\bf B}^2}{8\pi}}$ for the total pressure, where $p$ is the hydrodynamic pressure, $\rho$ density,
$\bf B$ is the magnetic field, $\bf g$  the acceleration due to gravity, $S$ the entropy, so that $-{\bf g} S$ is the buoyancy force;
$\displaystyle{
\rho_0 \mathcal{{\bf F}}_{mag}= \frac{({\bf B}\cdot\grad){\bf B}}{4\pi}
-\left(\frac{\grad \rho_0}{\rho_0}\right) \frac{{\bf B}^2}{8\pi} }$ the non-gradient part of the magnetic force in density stratified fluid, where the first turm stands for magnetic stress while the second term for magnetic buoyancy;
$\displaystyle{
 \mathcal{{\bf F}}_{hd} = {\bf u} \times {\bf w} }$ local Coriolis force from nonlinear local fluid motion,
and $\rho_0 \displaystyle{
\mathcal{{\bf F}}_{visc} = \nu \rho_0 {\left[
\nabla^2{\bf u} -\frac{2}{3}\nabla({\rm div}{\bf u}) \right]}
}$ the viscous force, where $\nu$ is the molecular viscosity,
and $\rho_0 \displaystyle{
\mathcal{{\bf F}}_{cor} = 2 \rho_0 {\bf u} \times {\bf \Omega_\sun}
}$ the Coriolis force from solar global rotation.

In order to eliminate the terms containing gradients of potentials we calculate {\curl} of that equation (\ref{vandak}), to obtain the equation
for vorticity ${\bec w}$, and we are interested in the radial component $w_r$ only.
We assume for the rough estimate that the contribution from the
$(\curl \mathcal{{\bec F}}_{mag})_r$, $(\curl
\mathcal{{\bec F}}_{hd})_r$,
$(\curl \mathcal{{\bec F}}_{visc})_r$,
$(\curl \mathcal{{\bec F}}_{cor})_r$,
can be replaced by a relaxation term as $-w_r/\tau_D$, where
$\tau_D$ has a meaning of the sunspot twisting time.

Under these assumptions the radial component of
the equation for the vorticity in the spherical coordinates reads
\vbox{
\begin{equation}
\frac{w_r}{\tau_D}
= 2(\curl[\bec u \times \bec \Omega])_r
=2 \left[
(\bec \Omega \cdot \bec \nabla)u_r
-\Omega_r {\rm div}{\bec u}
\right]
$$
$$=
2 \Omega
\left(
\cos{\theta}\frac{d u_r}{dr}
- \sin{\theta} \frac{1}{r} \frac{d u_r}{d\theta}
- \cos{\theta}{{\rm div}\,}{\bf u}
\right)
$$
$$=
-2 \Omega
\left[
\cos{\theta}
\left(
\frac{u_r}{H_\rho}
- \frac{d u_r}{dr}
\right)
+ \sin{\theta}
\frac{1}{r} \frac{d u_r}{d\theta}
\right]
\,,
\end{equation}
}
\noindent
where $\Omega$  is the solar angular rotation
approximately corresponding to Carrington rotation with the siderial rotation period of approximately 25 days.

The radial derivative of the vertical convective velocity can be estimated as
$\displaystyle{\frac{\partial u_r}{\partial r}}
\approx
-\displaystyle{\frac{u_r}{H_\rho}}$.
Here the negative sign reflects the effect of slow-down the velocity in the rising flux tubes.

\section{Estimation of the ratio of the toroidal and poloidal fields in dynamo models}

In order to estimate the ratio of the toroidal and poloidal fields we refer to the formalism of the paper by Kleeorin et al. (1995), especially their equation (3) and below.
Their equations are the non-dimensional $\alpha\Omega$-dynamo system for non-linear evolution of the poloidal (azimuthal component of the vector potential) $A$ and toroidal (azimuthal component of the magnetic field vector) $B$ fields.

The parameter $D=R_\alpha R_\Omega$ is the dimensionless dynamo number,
characterising the intensity of dynamo action that is defined using the
typical values of functions $\alpha$,
$\Omega$, and $\eta_T$, where $R_\alpha$ is the dimensionless number characterising the efficiency of the $\alpha-$~effect, and $R_\Omega$ is the dimensionless number characterising the differential rotation with respect to turbulent diffusivity $\eta_T$.

In the linear problem if the $\alpha$-coefficient is of the order unity, ratio $A/B$ is of order of $1/\sqrt{|D|}$.
Typical values of the dynamo number in developed nonlinear regime $D$ usually exceed the
critical value $ D_{cr}$ by factor $3-10$, and the range of $R_\alpha$ is typically of order $1-3$.
In the non-linear evolution, the effective $\alpha-$~coefficient is reduced by the order of $\xi=D_{\rm crit}/D$, where $D_{\rm crit}$ is the threshold value of the dynamo number for generation of a marginally unstable mode.
Thus, in the nonlinear regime the ratio $A/B$ becomes of order of $\sqrt{\xi /|D|} =\sqrt{|D_{\rm crit}|}/|D|$.
We can estimate the values of the dynamo number, e.g., using some simple 1D dynamo models reproducing basic regularities and irregularities of the solar cycle, see Kleeorin et al. (2016). They result in $D_{\rm crit}\approx -2\cdot10^3$ and $D\approx -8\cdot10^3$, therefore, the ratio is $A/B\approx 6\cdot10^{-3}$.


\end{document}